\documentclass[manuscript]{aastex}
\usepackage[caption=false]{subfig}
\usepackage{graphicx}

\shorttitle{Electron-Muon identification}
\shortauthors{Iori et al.}

\begin{document}


\title{Electron-muon identification by atmospheric shower and electron beam in a new concept of an EAS detector}

\author{M. Iori}
\affil{``Sapienza'' Univeristy of Rome, Piazzale A. Moro 5, Rome 00185, IT}
\email{Maurizio.Iori@roma1.infn.it}

\author{H. Denizli and A. Yilmaz}
\affil{Abant Izzet Baysal University, Bolu 14280,TR}

\author{F. Ferrarotto}
\affil{Istituto Nazionale di Fisica Nucleare (INFN) - Sezione di Roma, Rome 00185, IT}

\and

\author{J. Russ}
\affil{Carnegie-Mellon University, Pittsburgh, PA 15213, USA}



\begin{abstract}
We present results demonstrating the time resolution and $\mu$/e separation
capabilities with a new concept of an EAS detector capable for measurements of 
cosmic rays arriving with large zenith angles. This kind of detector has been 
designed to be a part of a large area (several square kilometers) surface 
array designed to measure Ultra High Energy (10-200 PeV) $\tau$ neutrinos using the Earth-skimming technique. 
A criteria  to identify electron-gammas is also shown and the particle identification capability is tested by measurements in coincidence with the KASKADE-GRANDE experiment in Karlsruhe, Germany.

\end{abstract}





\section{Introduction}

In the Ultra High Energy Cosmic Rays (UHECR) large variety of 2d (surface) or 3d (volumetric) detector arrays are constructed with different detection techniques, Cherenkov \citep{tunka_cherenkov,hegra_cherenkov}, air fluorescence \citep{fluoresrence} and radio waves \citep{radiowaves}). These experiments are mostly concentrate on detection of cosmic ray shower particles moving downward and the timing information is  used to obtain shower angular information; none of the present detectors uses precision time measurements to discriminate upward or downward  moving particles by Time Of Flight (TOF) and $\mu$/e separation. TOF discrimination is useful for experiments that seek to detect cosmic rays or cosmic neutrino interactions at zenith angles greater than $90^o$, $\mu$/e separation to improve the signature.
In this paper we show the results of a test on $\mu$/e separation, made in correlation with  KASCADE-GRANDE Experiment (KGE),  with a prototype module planned for deployment in an array capable of measuring large zenith angle cosmic rays as well as detecting the signature of Ultra High Energy $\tau$ neutrino interactions using the Earth 'skimming strategy' \citep{beacom,fargion99,fargion02,fargion,feng,zas}. Preliminary analysis of part of the data shown in this paper was presented in several conferences \citep{iori1,iori2}. The strategy described for $\mu$/e separation can be used also for future large array like LHAASO project if the TOF measurement is not implemented.

\subsection{Description of the module}
To detect single particle from a shower and determine its direction we have designed a detecor using two pairs of $20~\times~20~cm^2$ and 1.4 cm thick scintillator plates separated by 160 cm and  named tower. A single tower has a geometrical acceptance of $25.0~cm^2 sr$ and its zenith angle  range  $\pm~7.5^o$. In order to augment the acceptance of the detector, important to detect low intensity fluxes  like UHE neutrino flux, we put two towers  with their axes parallel and separated of 60 cm. This layout improves the acceptance of $\pm~20^o$ along the azimuthal angle.  Each tile is read by one low voltage, high time resolution R5783 Hamamatsu 
photomultiplier (PMT), extensively used in the CDF muon detector \citep{cdf}.
The PMT is connected to the scintillating tile and embedded in a PVC box.
The module in composed by four boxes attached to a metallic frame.
 The excellent PMT time resolution ($\approx 400~ps$) provides a good TOF measurements. The Cockroft-Walton use in PMT to generate high voltage
opens the possibility to power the system using a renewable energy power 
source like a solar panel or a wind turbine - an important feature for an 
elementary module in a large area array.  

With a TOF resolution of the order of $1~ns$ it is possible, when we measure
large zenith angle shower, to reject the vertical air showers without need of 
any shielding.  Hence we can select upward and downward particles passing through the detector with negligible intrinsic contamination. The module was setup vertically in the area of KASCADE-GRANDE array to perform the measurement parameters discussed in this paper.

At present the DAQ is based on waveform sampling, using a MATACQ system. This 
digitizes the scintillator waveform at $2GS/s$, covering a $2.5 ~\mu s$ window 
\citep{Matacq}. The MATACQ is triggered by an external signal that defines 
the direction of the track. The time of flight for determining whether the
track is moving up or down is refined offline, is obtained using an algorithm based on the linear fit of the front of photomoltiplier signal.

\subsection{The performances of the working prototype}

The performance of this module was initially tested at the High Altitude 
Research Station Jungfraujoch (HFSJG), located in Switzerland
at $\approx 3600~m$. Details of the first prototype are given in \citep{Ricap}. A second prototype was installed there in the summer of 2009 to test the latest electronics board \citep{iori1}. 
The module has shown a good upward/downward discrimination capability in 
all our tests. 
To reach the necessary upward-downward TOF discrimination the light 
collection technique has been optimized, focusing on the time resolution at the expense
of energy resolution. The definition of a vertical MIP is made by calibration 
on vertical downward cosmic rays, to set proper charge cuts to obtain a good 
time resolution. Because of the position variation of light collection
efficiency, this slightly reduces the effective area per \emph{tower}, but the
effect is small.
The time resolution we measure ($\approx 1.2~ns$), as shown in Fig. 2, is comparable to the PMT 
transit time spread. This is achieved by avoiding any reflections in the light collection process. 
To improve the time resolution we have connnected the $1~cm^{2}$ PMT window directly to the scintillator plate by a silicone rubber pad. By this configuration the first light arriving at PMT window dominates the leading edge of the signal.

\section{ Electron-gamma and muon identification}
\subsection{Electron-gamma}
Electron identification is obtained using a layer of lead with optimized 
thickness able to produce an  electromagnetic shower in front of a tile. 
The module in the test made  at KGE was set up vertically and a layer of 1.5 cm lead, corresponding to $3X_{0}$, covered the bottom tile (B). By using TOF we select downwarding vertically or/and diagonally tracks crossing the top (T)  and B tile of same tower or different towers, respectively. 

\begin{figure}
\begin{center}
\includegraphics[scale=0.5]{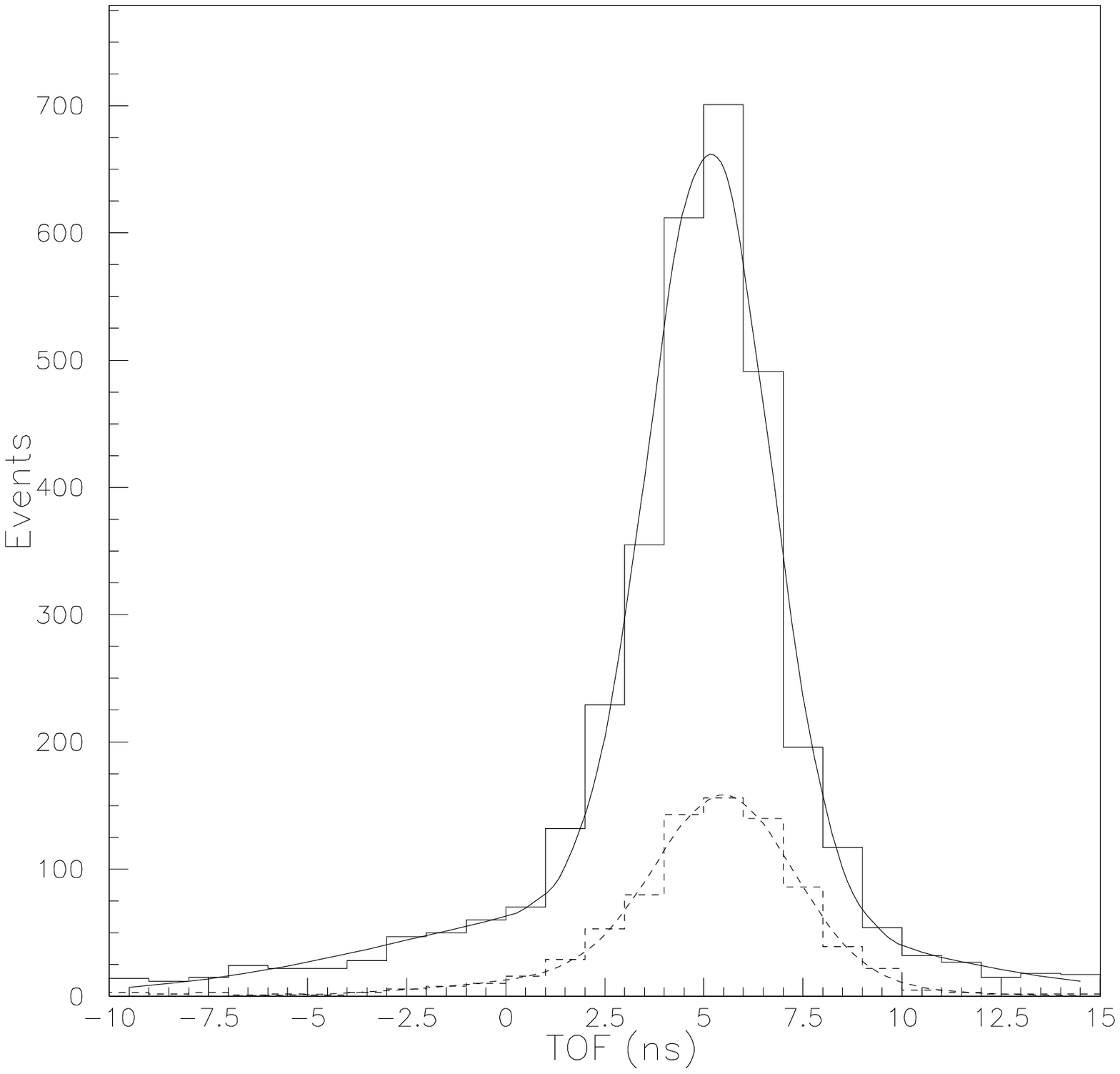}
\qquad 
\caption{\label{fig:tof} The distribution of time of flight between the two 20 $\times$ 20 $cm^{2}$ tiles $160~cm$ apart, for downward vertical cosmic ray events (solid line) and diagonal (broken line) in run with a 1.5 cm lead of layer. The towers are apart 60 cm. In this figure we performed two gaussian fit to cover a tail as well. For vertical and diagonal tracks sigma is about 1.3 and 1.7, respectively. That shows within 2$\sigma$ the TOF is 5 $\pm$ 3 ns.}
\end{center}
\end{figure}

An algorithm that extrapolates the front of the signal to zero voltage after applying 6 mV threshold cut (the mean amplitude of signal is 60 mV after cable attenuation) is used to find time for the arrival, $t_{0}$, of the track. Then TOF was evaluated simply by taking the difference between $t_{0}$ of the bottom and top tiles. Fig. \ref{fig:tof} shows the TOF distributions for vertical and diagonal tracks. 

For the analysis we used two data sets. One with lead layer inserted at the top of the bottom tile with 5792 \emph{good} tracks recostructed, the other with no lead layer with 2520 \emph{good} tracks. The term \emph{good} means the TOF of track was in the interval 5 $\pm$ 3 ns that corresponds to 2 sigma. The reason why we used lead layer in front of the bottom tile is to produce more light at the B tile due to electromagnetic processes such as pair production or Compton effect for electrons or gammas,($\gamma$). Lead layer will show no differences in T and B tiles for muons ($\mu$). 
A good variable to separate the electromagnetic component from $\mu$ is the ratio (R) of the charge in B tile ($ q_{B} $) to the charge deposited in T tile, ($ q_{T} $). 

 By GEANT4 we have simulated normally incident uniformed flux of $\gamma$, electrons and $\mu$'s with different momentum (50, 100, 150, 200 and 500 MeV) passing through the scintillating T tile and with and without lead layer in front of B tile. Then we have evaluated the ratio R obtained as energy deposited in the scintillator in B tile divided by in the T tile. 
Evaluating the ratio R we applied 1 MeV energy cut for the T and B tiles corresponding to the energy loss in the scintillating tiles.
 The ratio R of $\gamma$'s, electrons and $\mu$'s at different energies  without lead (upper row) and with lead (lower row) is shown in \citep{iori3} . The $\gamma$ and electron R distributions indicates an evident difference of the slope at R$>$ 2 between the run with lead because of more energy deposited in the B tile. 
In the case without lead the electron R distribution is clustered near unity close to 1, quite similar to the $\mu$'s. 
The muon R value is always close to 1 except for 50 MeV $\mu$'s where the Coulomb effect in the lead layer produces electrons that spread the R distribution. 
The 500 MeV $\gamma$'s and electrons have a broad R distribution because of the relativistic rise in the energy loss.  
Different gain and light collection might influence the measurement of R distribution. To correct this effect we have selected a sample of tracks with a single peak signal in both tiles. 
We expect this sample will contain mostly $\mu$  tracks will deposit the same energy in both tiles with Landau fluctuation. 
Therefore this sample should produce a narrow charge distribution close to a mean value of total charge collected by photomultiplier.

 Comparing the mean values of charge distribution between T and B tiles we calculate the ratio correction factor, $k$.
The correction factor was calculated by comparing the mean value of the charge distribution of the T tile and B tile. It results of each tower, $k$ to be $ 0.89 \pm 0.09$ and $0.82 \pm 0.08$ respectively. To verify the presence of lead we have compared the R and the inverse ratio $R^{*} = {q_{T}\over q_{B}}$ distributions. We know the R and R* distribution when the lead is absent should be almost equal, the symmetry can be broken by downward moving tracks that interact in the T tile.

\begin{figure}
\begin{center}
\includegraphics [width=8.cm,totalheight=12.cm]{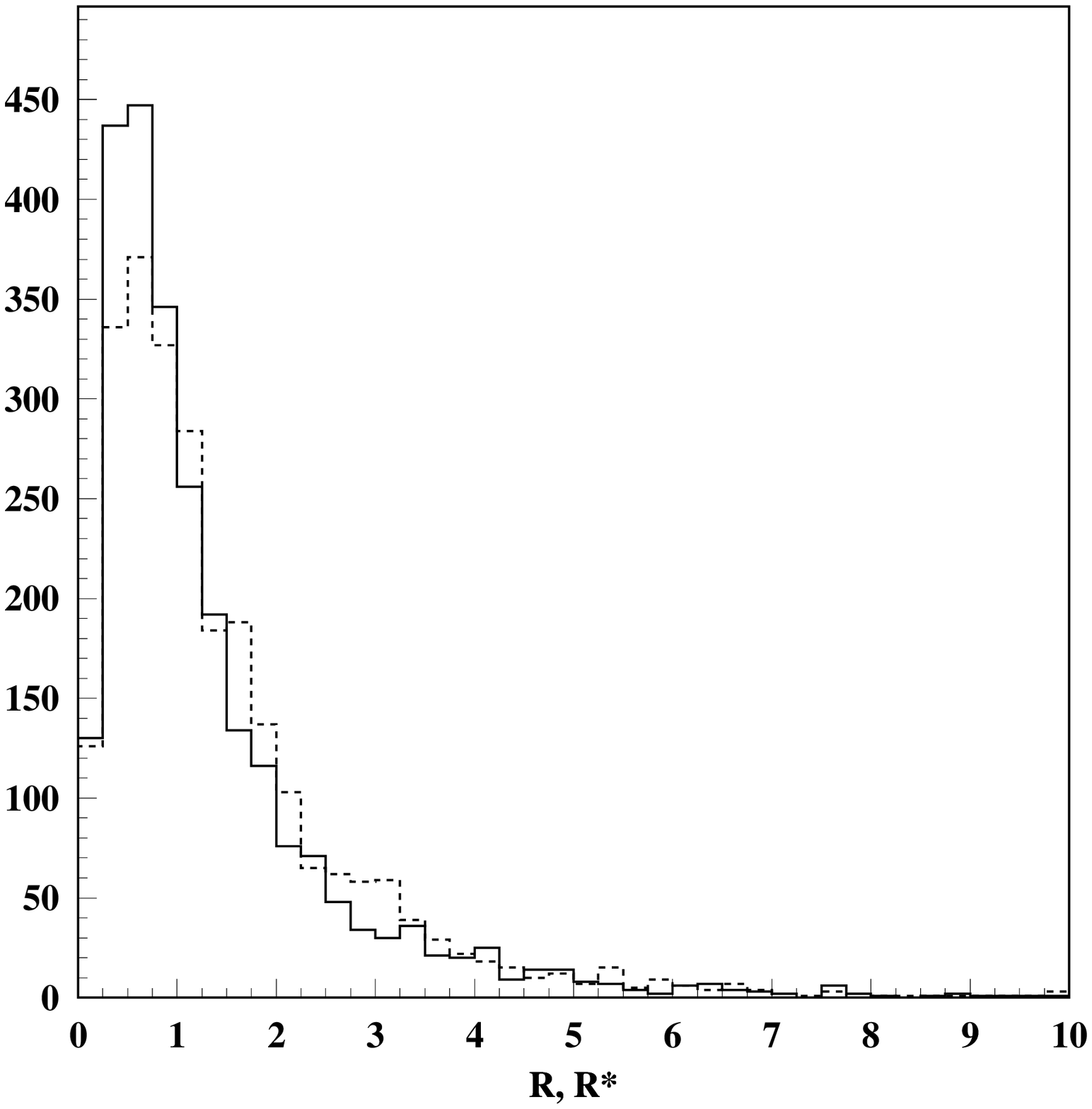}
\includegraphics [width=8.cm,totalheight=12.cm]{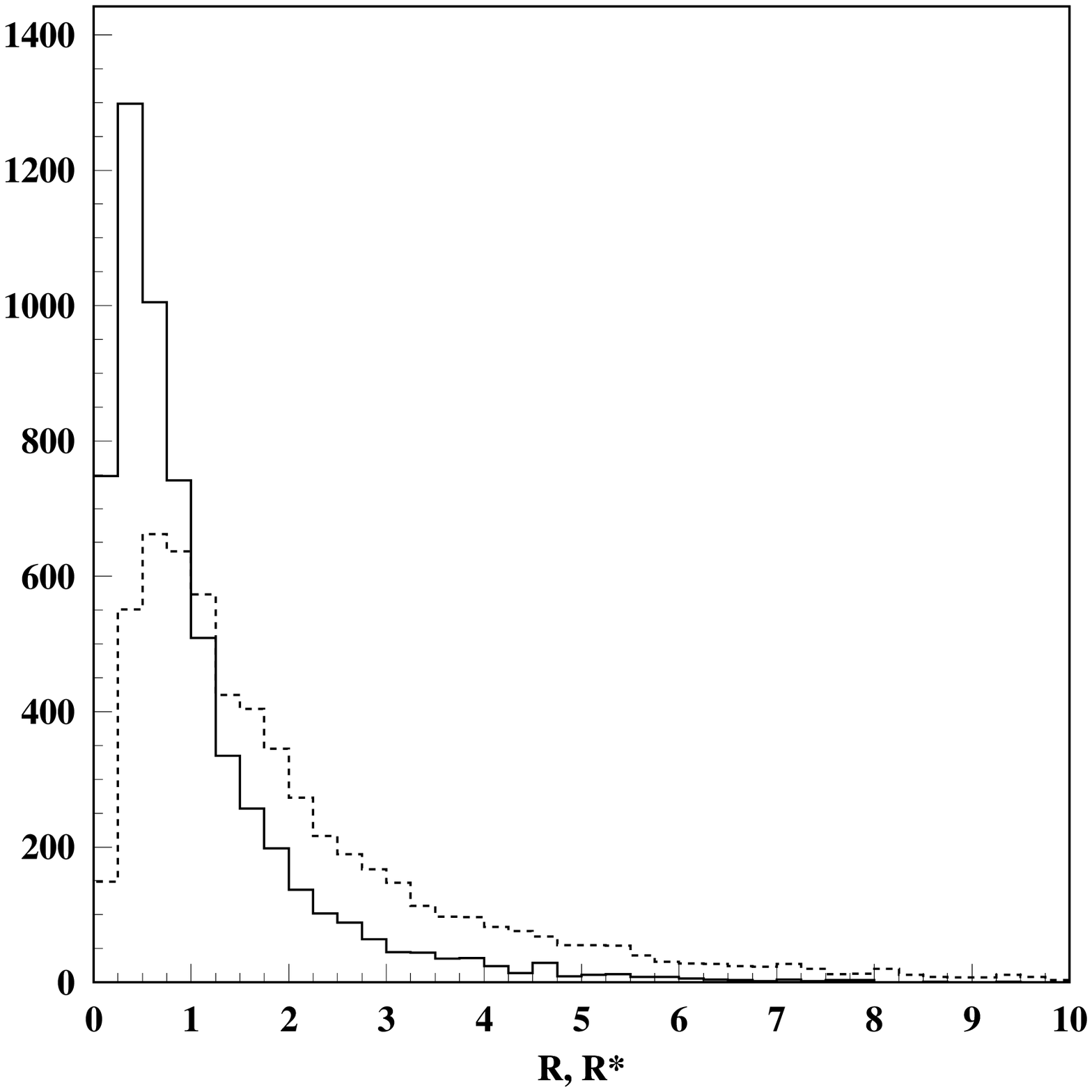}
\caption{\label{fig:RNL} The corrected ratio $R^{*}$ (solid) and $R$ (dot) of the \emph{good} tracks from run without (left) and with lead (right). The small excess of events in R distributions above R=2 comes from more interactions in the B tiles that leads to increase the released charge in the B tile.}
\end{center}
\end{figure}

Fig.\ref{fig:RNL} shows the corrected ratio $R^{*}$ (solid) and $R$ (dotted)
of the \emph{good} tracks in the run without (left) and with lead (right). 
The slope is the same and the loss of events in the $R^{*}$ distribution (solid) below R equal 1 is due to the interactions in the T tiles ${q_{T}< q_{B}}$ as verified by GEANT4.
Fig.\ref{fig:RNL} (right) indicates an obvious increase on the number of events at R$>2$. This increment is related to a production of the electromagnetic component in lead layer and detected by the B tile. 

In order to quantify the contributions coming from electromagnetic components in R distributions, we performed MINUIT fit to experimental data using R distributions obtained from GEANT4 simulations done for electrons, $\gamma$'s and $\mu$'s with energies starting from 50 MeV  up to 500 MeV. Density of particles of  a shower are  generated in the detector acceptance as a function of momentum and used to weight the R  distribution. Density strongly depends on the distance from the shower center. At 25 m from center of core the  density drops to approximately 10 particles/$m^{2}$ for both electrons and $\gamma$'s reducing the probability to detect vertical tracks .

Using primary particle as proton, 100 vertical showers are created. Results show that 57\% of $\gamma$'s with momentum less than 150 MeV,  16\% between 150 MeV and 250 MeV are inside a distance of 25 m from the core in the detector acceptance. The rest of them (27\%) have higher momentum. 61\% of electrons have momentum less than 250 MeV. 74\% of $\mu$'s have momentum greater than 500 MeV, \citep{corsika}.

The predicted number of events for electron, $\gamma$ and $\mu$ in the bin, $i$, used to fit the experimental distribution, is

$$F(p_{j},R_{i})_{e,\gamma, \mu}= N_{D}(w(p_{j})f_{p_{j}}(R_{i}))_{e,\gamma, \mu}$$

where $N_{D}$ is the total number in the data sample, $w(p_{j})$ is the normalized density of tracks around the detector for three different intervals of momentum with average, in an interval of $\pm$ 50 MeV, 100 MeV, 200 MeV and 500 MeV for electrons, 100, 200 MeV and 500 MeV for $\gamma$'s and 200 MeV and 500 MeV for $\mu$'s inside a radius of 25 m, $f_{p_{j}}(R_{i})$ is the fraction of simulated tracks by GEANT4 in each bin as function of the momentum.

The best result corresponds to a detected track sample of 86.7 $\pm$ 1.6\% electrons, 5.4 $\pm$ 0.4 \% $\gamma$'s and 7.9 $\pm$ 0.8  \% of $\mu$'s.

We also generated showers at 5$\times 10{^7}$ GeV using primary particle as Carbon and Iron. Assuming about 20$\%$ of Iron in the cosmic ray flux \citep{components}, we have found the fit results  within 2$\sigma$. Changing the range of momentum distribution used in the fit and the size of the core from 25 meters to 100 meters the fit results are also within 2$\sigma$.

At Fermilab Meson Test Beam Facility we took data using a 500 MeV electron beam.  Electrons were identified by a gas Cherenkov counter.  They traversed two scintillator tiles spaced 6 cm apart, with 1.5 cm of lead covering the downstream tile.  We took data in samples over the entire tile area.  The left 10\% of the downstream scintillator was not covered by the lead.  We used events in this region to normalize the pulse height distributions between the two scintillator tiles.  The measured pulse height ratio for the counter after the lead to that before was 3.56 $\pm$ 0.14 in good agreement with the GEANT4 prediction.  A simple ray-tracing program  predicts the variation of the ratio across the face of the scintillator and the comparison of the test beam to the data is in agreement of 10\%.

\subsection {KASCADE-GRANDE validation}
 
After we have isolated by fitting the percentage of electrons and $\mu$'s in the sample of tracks selected by TOF we considered the particle identification provided by KGE operating in coincidence in a gate of 600 ms with our detector to validate the criteria.
The KASCADE array has large shower reconstruction efficiency for incoming 
cosmic rays having zenith angles from vertical to 40 degrees. 

For each event the KGE predicts by the lateral shower distribution the $\mu$ and electron density per square meter as distance from the core. The electrons and $\mu$'s identified by KGE have minimum energy of 5 MeV and 230 MeV respectively and no direction information. 
Our module selects tracks from the KGE data with zenith angles in the range $\pm$ 7.5 degrees. 
The KGE trigger rate for all 16 sectors is about 3.0 Hz. The events are recorded only for the KGE events with a shower core in sector 14, where our module has been installed, with a rate of 0.18 Hz. 56$\%$ of triggers have a reconstructed shower into the KGE sector.
In the triggered data the rate at which our module shows at least
one hit tile is 0.015 Hz, or 1 hit for each 12 KGE triggers. Of these single-hit events 3$\%$ have a reconstructed track in the tower despite the small sampling area. By using CORSIKA simulation of vertical showers as reconstructed by KGE we have found a good agreement with the percentage of the reconstructed tracks. 
The sample of the good tracks for a 1.5 cm lead absorber which were selected by TOF are verified by the KGE data to be a $\mu$ or electron by selecting KGE 
events in the sensitive area where there is one and more shower track, either electron or $\mu$. To validate the R distribution in our module we normalized the particle density provided by KGE to the acceptance of our module. 
 
\begin{figure}
\begin{center}
 \subfloat[electrons]{{\includegraphics[scale=0.4]{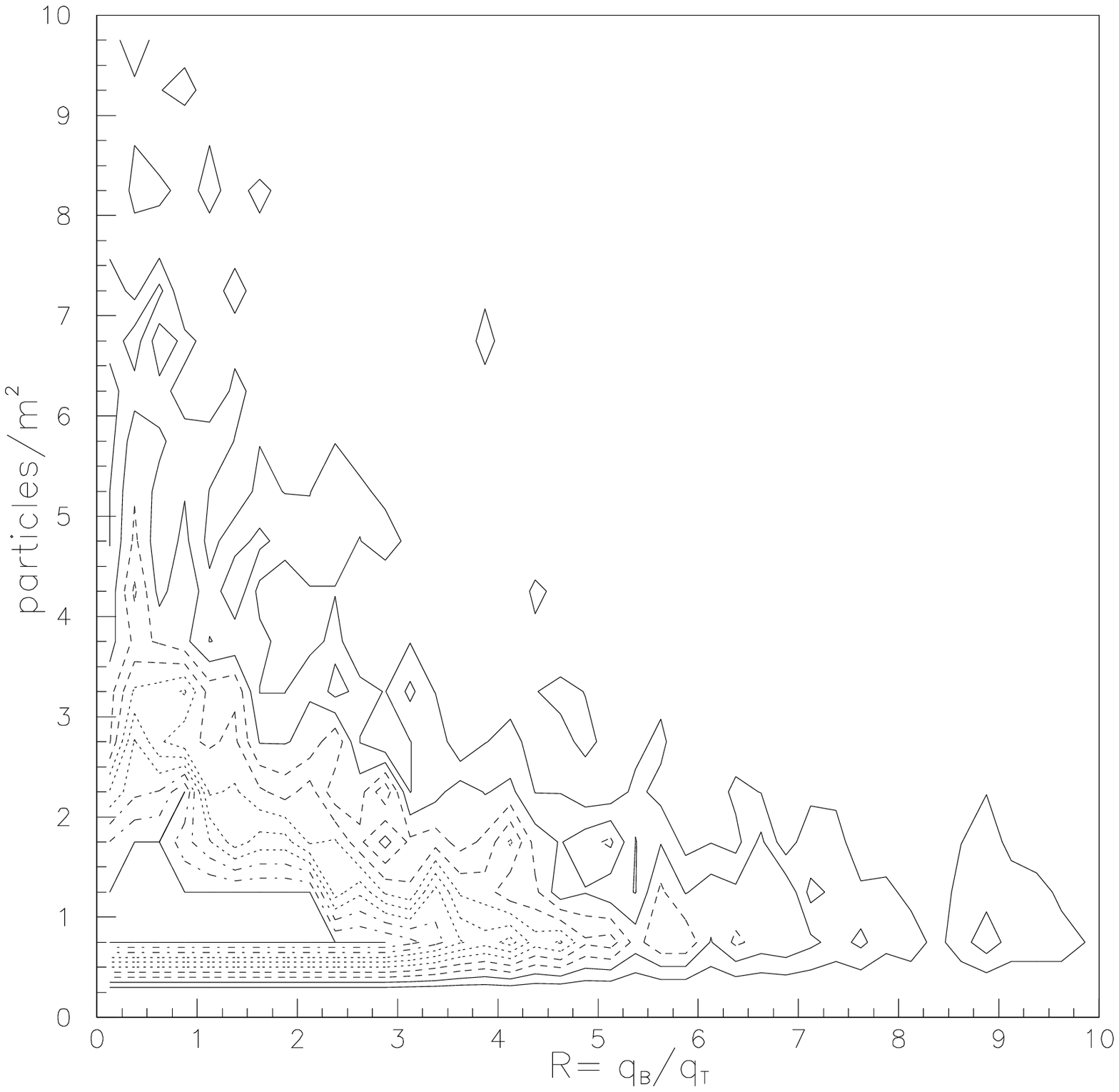} }}%
 \qquad 
\subfloat[muons]{{\includegraphics[scale=0.4]{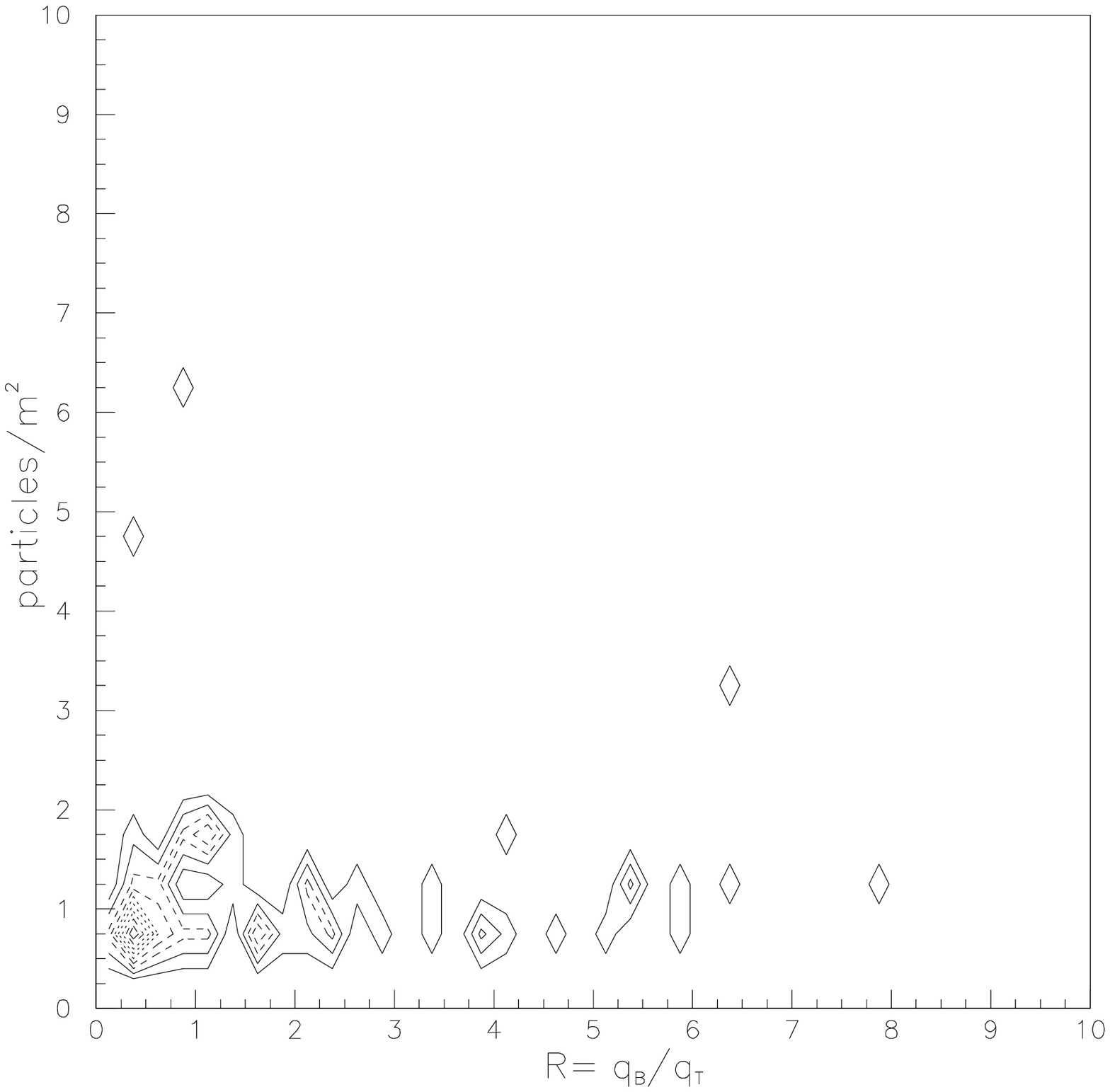} }}%
\caption{\label{fig:valid} KGE track density on $m^{2}$ versus  R distribution for electrons (a) and muons tracks (b) reconstruceted in the detector and validated by KGE.}
\end{center}
\end{figure}

Figure \ref{fig:valid} shows the electron (a) and $\mu$ (b) density (particles/$m^2$) versus R obtained by the interpolation of the electron/$\mu$ multiplicity from the  KGE stations closer to our detector. The Figure \ref{fig:valid}a results requiring $\mu$ density less than $10^{-3}$ particles/$m^2$. The Figure \ref{fig:valid}b requires at least a $\mu$ predicted by KGE in the detector area without any cut on electron density. The $\mu$'s are mainly concentrated at R=1.  The events with R$>$2 can be interpretated, due to the contribution of electrons with a large momentum ($>$ 500 MeV) or $\mu$'s with a low momentum ($<$ 50 MeV) as predicted by GEANT4 studies on MonteCarlo \citep{iori3}. To evaluate this contamination we used the sample of 100 vertical showers generated by CORSIKA at 5$\times 10{^7}$ GeV (p,C and Fe) and selected the muons with electrons in a $m^2$ within 25 m. We have found 36.3 $\%$ of the electrons in the shower core have momentum higher than 500 MeV in presence of a muon and 80$\%$ are within 25 m away from the center.That explains the tail at R$>2$ is due to hard electrons present in the detector acceptance when there is a $\mu$.
The aim of this prototype is to measure tracks at large zenith angle so we performed a test where the detector was put horizontally to evaluate the noise from two vertical tracks shifted in time and with TOF of about 5 ns. The probability obtained counting this category of  events with the TOF within $\pm$(3:7) ns over all events with signals on both tiles results to be $10^{-2}$ for a single tower and a rate of $10^{-6}$ Hz.

\section{Conclusions}

By using the KASCADE-GRANDE trigger we have tested a method to select low momentum electron-$\mu$ on a prototype designed to measure horizontal cosmic ray flux. By optimizing the thickness of a lead layer located on the surface of the B scintillating tile we find that the tracks deposite more energy in B tile are electron with momentum of about 200 MeV. The analysis suggests when we require R is bigger than 2 we select electrons with no contamination of $\mu$'s.If we discard the TOF measurement, a station made by a layer of lead between the T and B large tiles can be used in a new generation of experiments like LHAASO project.

\acknowledgments
\section{Acknowledgements:}
The Scientific and Technical Research Council of Turkey (TUBITAK) is acknowledged for granting Ali YILMAZ a International Doctoral Research study in the framework of TUBITAK-BIDEB 2214 grant. The authors thank to the KASCADE-GRANDE Collaboration for hosting our detector at Forschungszentrum Karlsruhe/Karlsruhe Institute of Technology, especially to Harald Schieler, Andreas Weindl, Victor de Souza, and Bernd Hofmann for their technical help.We thank Mauro Ciaccafava for technical work and Giacomo Chiodi, Riccardo Lunadei of INFN-Labe. We thank the National Science Foundation  and the INFN for financial aid. Authors also thank to ERASMUS program for opportunity to establish a collaboration between University of Roma and Abant Izzet Baysal University.

\clearpage

\clearpage

\end{document}